# Market-Oriented Cloud Computing:
# Vision, Hype, and Reality for Delivering IT Services as Computing Utilities

Rajkumar Buyya[1,2], Chee Shin Yeo[1], and Srikumar Venugopal[1]

[1] **Gri**d Computing and **D**istributed **S**ystems (GRIDS) Laboratory
Department of Computer Science and Software Engineering
The University of Melbourne, Australia
Email: {raj, csyeo, srikumar}@csse.unimelb.edu.au

[2] Manjrasoft Pty Ltd, Melbourne, Australia

## Abstract

This keynote paper: presents a 21$^{st}$ century vision of computing; identifies various computing paradigms promising to deliver the vision of computing utilities; defines Cloud computing and provides the architecture for creating market-oriented Clouds by leveraging technologies such as VMs; provides thoughts on market-based resource management strategies that encompass both customer-driven service management and computational risk management to sustain SLA-oriented resource allocation; presents some representative Cloud platforms especially those developed in industries along with our current work towards realising market-oriented resource allocation of Clouds by leveraging the 3$^{rd}$ generation Aneka enterprise Grid technology; reveals our early thoughts on interconnecting Clouds for dynamically creating an atmospheric computing environment along with pointers to future community research; and concludes with the need for convergence of competing IT paradigms for delivering our 21$^{st}$ century vision.

## 1. Introduction

With the advancement of the modern human society, basic essential services are commonly provided such that everyone can easily obtain access to them. Today, utility services, such as water, electricity, gas, and telephony are deemed necessary for fulfilling daily life routines. These utility services are accessed so frequently that they need to be available whenever the consumer requires them at any time. Consumers are then able to pay service providers based on their usage of these utility services.

In 1969, Leonard Kleinrock [1], one of the chief scientists of the original Advanced Research Projects Agency Network (ARPANET) project which seeded the Internet, said: "As of now, computer networks are still in their infancy, but as they grow up and become sophisticated, we will probably see the spread of '*computer utilities*' which, like present electric and telephone utilities, will service individual homes and offices across the country." This vision of the computing utility based on the service provisioning model anticipates the massive transformation of the entire computing industry in the 21$^{st}$ century whereby computing services will be readily available on demand, like other utility services available in today's society. Similarly, computing service users (consumers) need to pay providers only when they access computing services. In addition, consumers no longer need to invest heavily or encounter difficulties in building and maintaining complex IT infrastructure.

Software practitioners are facing numerous new challenges toward creating software for millions of consumers to use as a service rather than to run on their individual computers. Over the years, new computing paradigms have been proposed and adopted, with the emergence of technological advances such as multi-core processors and networked computing environments, to edge closer toward achieving this grand vision. As shown in Figure 1, these new computing paradigms include cluster computing, Grid computing, P2P computing, service computing, market-oriented computing, and most recently Cloud computing. All these paradigms promise to provide certain attributes or capabilities in order to realize the possibly 1 trillion dollars worth of the utility/pervasive computing industry as quoted by Sun Microsystems co-founder Bill Joy [2]. Computing services need to be highly reliable, scalable, and autonomic to support ubiquitous access, dynamic discovery and composability. In particular, consumers can determine the required service level through Quality of Service (QoS) parameters and Service Level Agreements (SLAs). Of all these computing paradigms, the two most promising ones appear to be Grid computing and Cloud computing.

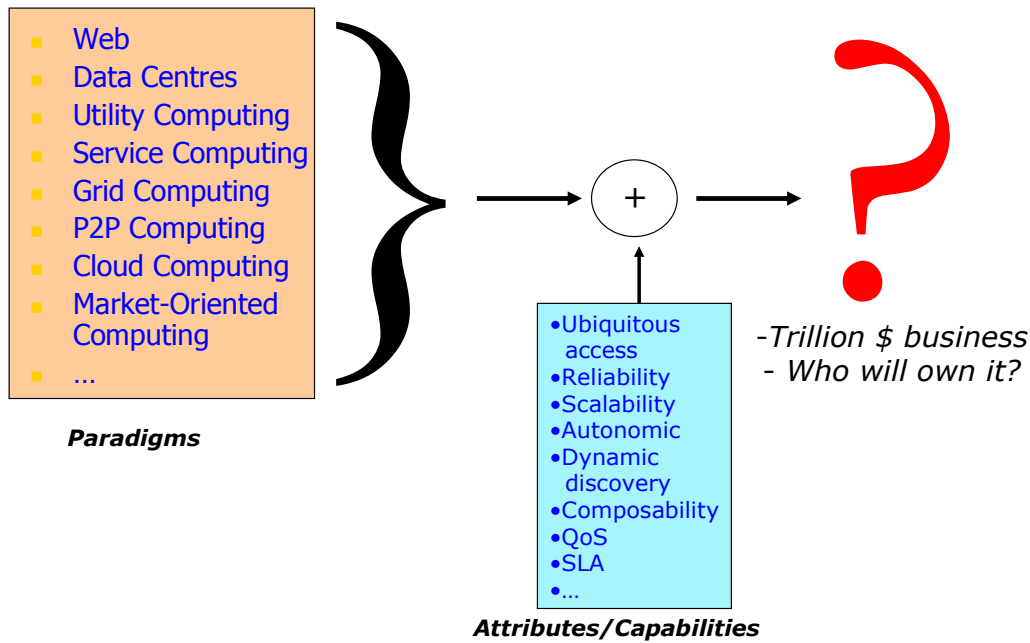

**Figure 1:** Various paradigms promising to deliver IT as services.

A Grid [3] enables the sharing, selection, and aggregation of a wide variety of geographically distributed resources including supercomputers, storage systems, data sources, and specialized devices owned by different organizations for solving large-scale resource-intensive problems in science, engineering, and commerce. Inspired by the electrical power Grid's pervasiveness, ease of use, and reliability [4], the motivation of Grid computing was initially driven by large-scale, resource (computational and data)-intensive scientific applications that required more resources than a single computer (PC, workstation, supercomputer, or cluster) could have provided in a single administrative domain. Due to its potential to make impact on the 21st century as much as the electric power Grid did on the 20th century, Grid computing has been hailed as the next revolution after the Internet and the Web.

Today, the latest paradigm to emerge is that of Cloud computing [5] which promises reliable services delivered through next-generation data centers that are built on compute and storage virtualization technologies. Consumers will be able to access applications and data from a "Cloud" anywhere in the world on demand. In other words, the Cloud appears to be a single point of access for all the computing needs of consumers. The consumers are assured that the Cloud infrastructure is very robust and will always be available at any time.

### 1.1 Definition and Trends

A number of computing researchers and practitioners have attempted to define Clouds in various ways [6]. Based on our observation of the essence of what Clouds are promising to be, we propose the following definition:

- "A Cloud is a type of parallel and distributed system consisting of a collection of inter-connected and virtualised computers that are dynamically provisioned and presented as one or more unified computing resources based on service-level agreements established through negotiation between the service provider and consumers."

At a cursory glance, Clouds appear to be a combination of clusters and Grids. However, this is not the case. Clouds are clearly next-generation data centers with nodes "virtualized" through hypervisor technologies such as VMs, dynamically "provisioned" on demand as a personalized resource collection to meet a specific service-level agreement, which is established through a "negotiation" and accessible as a composable service via "Web 2.0" technologies.

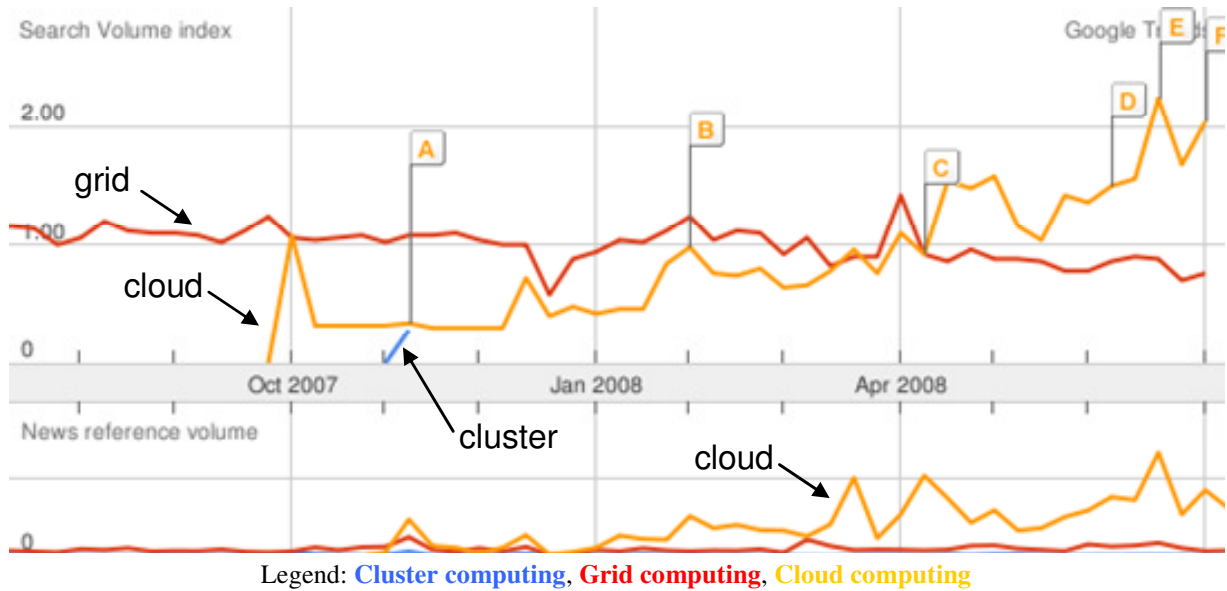

Legend: **Cluster computing**, **Grid computing**, **Cloud computing**

**Figure 2:** Google search trends for the last 12 months.

## 1.2 Web Search Trends

The popularity of different paradigms varies with time. The Web search popularity, as measured by the Google search trends during the last 12 months, for terms "cluster computing", "Grid computing", and "Cloud computing" is shown in Figure 2. From the Google trends, it can be observed that cluster computing was a popular term during 1990s, from early 2000 Grid computing become popular, and recently Cloud computing started gaining popularity.

Spot points in Figure 2 indicate the release of news related to Cloud computing as follows:

[A] IBM Introduces 'Blue Cloud' Computing, CIO Today - Nov 15 2007

[B] IBM, EU Launch RESERVOIR Research Initiative for Cloud Computing, IT News Online - Feb 7 2008

[C] Google and Salesforce.com in Cloud computing deal, Siliconrepublic.com - Apr 14 2008

[D] Demystifying Cloud Computing, Intelligent Enterprise - Jun 11 2008

[E] Yahoo realigns to support Cloud computing, 'core strategies', San Antonio Business Journal - Jun 27 2008

[F] Merrill Lynch Estimates "Cloud Computing" To Be $100 Billion Market, SYS-CON Media - Jul 8 2008

## 2. Market-Oriented Cloud Architecture

As consumers rely on Cloud providers to supply all their computing needs, they will require specific QoS to be maintained by their providers in order to meet their objectives and sustain their operations. Cloud providers will need to consider and meet different QoS parameters of each individual consumer as negotiated in specific SLAs. To achieve this, Cloud providers can no longer continue to deploy traditional system-centric resource management architecture that do not provide incentives for them to share their resources and still regard all service requests to be of equal importance. Instead, market-oriented resource management [7] is necessary to regulate the supply and demand of Cloud resources at market equilibrium, provide feedback in terms of economic incentives for both Cloud consumers and providers, and promote QoS-based resource allocation mechanisms that differentiate service requests based on their utility.

Figure 3 shows the high-level architecture for supporting market-oriented resource allocation in Data Centers and Clouds. There are basically four main entities involved:

- ***Users/Brokers***: Users or brokers acting on their behalf submit service requests from anywhere in the world to the Data Center and Cloud to be processed.

- ***SLA Resource Allocator***: The SLA Resource Allocator acts as the interface between the Data Center/Cloud service provider and

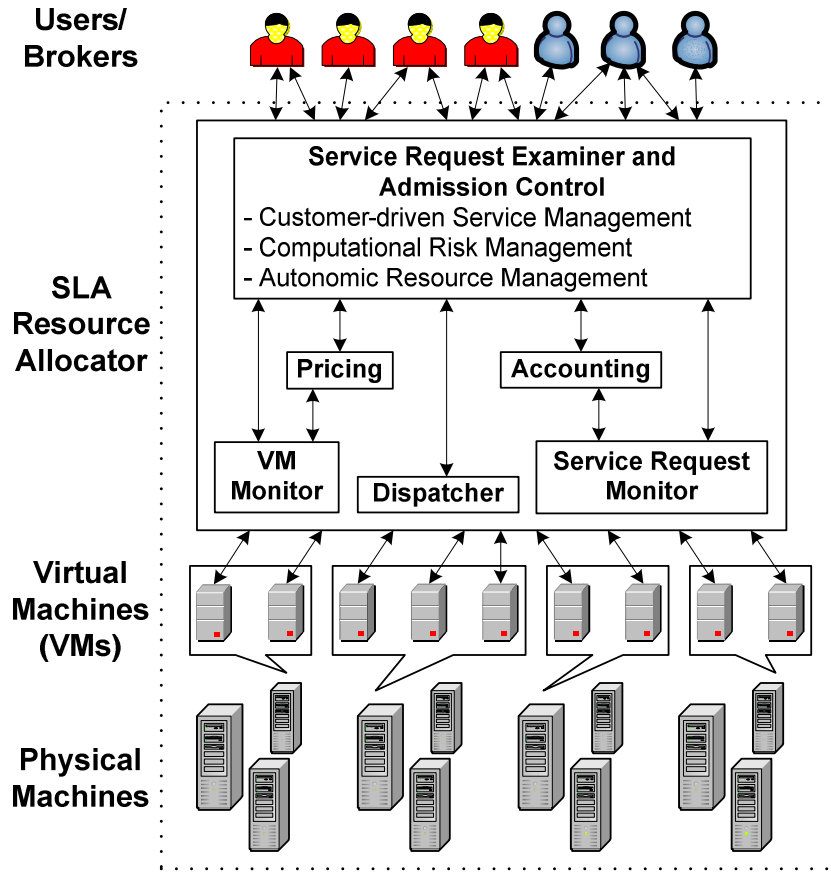

**Figure 3:** High-level market-oriented cloud architecture.

external users/brokers. It requires the interaction of the following mechanisms to support SLA-oriented resource management:

- o *Service Request Examiner and Admission Control*: When a service request is first submitted, the Service Request Examiner and Admission Control mechanism interprets the submitted request for QoS requirements before determining whether to accept or reject the request. Thus, it ensures that there is no overloading of resources whereby many service requests cannot be fulfilled successfully due to limited resources available. It also needs the latest status information regarding resource availability (from VM Monitor mechanism) and workload processing (from Service Request Monitor mechanism) in order to make resource allocation decisions effectively. Then, it assigns requests to VMs and determines resource entitlements for allocated VMs.

- o *Pricing*: The Pricing mechanism decides how service requests are charged. For instance, requests can be charged based on submission time (peak/off-peak), pricing rates (fixed/changing) or availability of resources (supply/demand). Pricing serves as a basis for managing the supply and demand of computing resources within the Data Center and facilitates in prioritizing resource allocations effectively.

- o *Accounting*: The Accounting mechanism maintains the actual usage of resources by requests so that the final cost can be computed and charged to the users. In addition, the maintained historical usage information can be utilized by the

Service Request Examiner and Admission Control mechanism to improve resource allocation decisions.
- o *VM Monitor*: The VM Monitor mechanism keeps track of the availability of VMs and their resource entitlements.
- o *Dispatcher*: The Dispatcher mechanism starts the execution of accepted service requests on allocated VMs.
- o *Service Request Monitor*: The Service Request Monitor mechanism keeps track of the execution progress of service requests.
- **VMs:** Multiple VMs can be started and stopped dynamically on a single physical machine to meet accepted service requests, hence providing maximum flexibility to configure various partitions of resources on the same physical machine to different specific requirements of service requests. In addition, multiple VMs can concurrently run applications based on different operating system environments on a single physical machine since every VM is completely isolated from one another on the same physical machine.
- **Physical Machines:** The Data Center comprises multiple computing servers that provide resources to meet service demands.

In the case of a Cloud as a commercial offering to enable crucial business operations of companies, there are critical QoS parameters to consider in a service request, such as time, cost, reliability and trust/security. In particular, QoS requirements cannot be static and need to be dynamically updated over time due to continuing changes in business operations and operating environments. In short, there should be greater importance on customers since they pay for accessing services in Clouds. In addition, the state-of-the-art in Cloud computing has no or limited support for dynamic negotiation of SLAs between participants and mechanisms for automatic allocation of resources to multiple competing requests. Recently, we have developed negotiation mechanisms based on alternate offers protocol for establishing SLAs [8]. These have high potential for their adoption in Cloud computing systems built using VMs.

Commercial offerings of market-oriented Clouds must be able to:

- support customer-driven service management based on customer profiles and requested service requirements,
- define computational risk management tactics to identify, assess, and manage risks involved in the execution of applications with regards to service requirements and customer needs,
- derive appropriate market-based resource management strategies that encompass both customer-driven service management and computational risk management to sustain SLA-oriented resource allocation,
- incorporate autonomic resource management models that effectively self-manage changes in service requirements to satisfy both new service demands and existing service obligations, and
- leverage VM technology to dynamically assign resource shares according to service requirements.

## 3. Emerging Cloud Platforms

Industry analysts have made bullish projections on how Cloud computing will transform the entire computing industry. According to a recent Merrill Lynch research note [9], Cloud computing is expected to be a "$160-billion addressable market opportunity, including $95-billion in business and productivity applications, and another $65-billion in online advertising". Another research study by Morgan Stanley [10] has also identified Cloud computing as one of the prominent technology trends. As the computing industry shifts toward providing Platform as a Service (PaaS) and Software as a Service (SaaS) for consumers and enterprises to access on demand regardless of time and location, there will be an increase in the number of Cloud platforms available. Recently, several academic and industrial organisations have started investigating and developing technologies and infrastructure for Cloud Computing. Academic efforts include Virtual Workspaces [11] and OpenNebula [12]. In this section, we compare six representative Cloud platforms with industrial linkages in Table 1.

Amazon Elastic Compute Cloud (EC2) [13] provides a virtual computing environment that enables a user to run Linux-based applications. The user can either create a new Amazon Machine Image (AMI) containing the applications, libraries, data and associated configuration settings, or select from a library of globally available AMIs. The user then needs to upload the created or selected AMIs to Amazon

**Table 1:** Comparison of some representative Cloud platforms.

| System / Property | Amazon Elastic Compute Cloud (EC2) | Google App Engine | Microsoft Live Mesh | Sun Network.com (Sun Grid) | GRIDS Lab Aneka |
|---|---|---|---|---|---|
| **Focus** | Infrastructure | Platform | Infrastructure | Infrastructure | Software Platform for enterprise Clouds |
| **Service Type** | Compute, Storage (Amazon S3) | Web application | Storage | Compute | Compute |
| **Virtualisation** | OS Level running on a Xen hypervisor | Application container | OS level | Job management system (Sun Grid Engine) | Resource Manager and Scheduler |
| **Dynamic Negotiation of QoS Parameters** | None | None | None | None | SLA-based Resource Reservation on Aneka side. |
| **User Access Interface** | Amazon EC2 Command-line Tools | Web-based Administration Console | Web-based Live Desktop and any devices with Live Mesh installed | Job submission scripts, Sun Grid Web portal | Workbench, Web-based portal |
| **Web APIs** | Yes | Yes | Unknown | Yes | Yes |
| **Value-added Service Providers** | Yes | No | No | Yes | No |
| **Programming Framework** | Customizable Linux-based Amazon Machine Image (AMI) | Python | Not applicable | Solaris OS, Java, C, C++, FORTRAN | APIs supporting different programming models in C# and other .Net supported languages |

Simple Storage Service (S3), before he can start, stop, and monitor instances of the uploaded AMIs. Amazon EC2 charges the user for the time when the instance is alive, while Amazon S3 charges for any data transfer (both upload and download).

Google App Engine [14] allows a user to run Web applications written using the Python programming language. Other than supporting the Python standard library, Google App Engine also supports Application Programming Interfaces (APIs) for the datastore, Google Accounts, URL fetch, image manipulation, and email services. Google App Engine also provides a Web-based Administration Console for the user to easily manage his running Web applications. Currently, Google App Engine is free to use with up to 500MB of storage and about 5 million page views per month.

Microsoft Live Mesh [15] aims to provide a centralized location for a user to store applications and data that can be accessed across required devices (such as computers and mobile phones) from anywhere in the world. The user is able to access the uploaded applications and data through a Web-based Live Dekstop or his own devices with Live Mesh software installed. Each user's Live Mesh is password-protected and authenticated via his Windows Live Login, while all file transfers are protected using Secure Socket Layers (SSL).

Sun network.com (Sun Grid) [16] enables the user to run Solaris OS, Java, C, C++, and FORTRAN based applications. First, the user has to build and debug his applications and runtime scripts in a local development environment that is configured to be similar to that on the Sun Grid. Then, he needs to create a bundled zip archive (containing all the related scripts, libraries, executable binaries and input data) and upload it to Sun Grid. Finally, he can execute and monitor the application using the Sun Grid Web portal or API.

After the completion of the application, the user will need to download the execution results to his local development environment for viewing.

GRIDS Lab Aneka [17], which is being commercialized through Manjrasoft, is a .NET-based service-oriented platform for constructing enterprise Grids. It is designed to support multiple application models, persistence and security solutions, and communication protocols such that the preferred selection can be changed at anytime without affecting an existing Aneka ecosystem. To create an enterprise Grid, the service provider only needs to start an instance of the configurable Aneka container hosting required services on each selected desktop computer. The purpose of the Aneka container is to initialize services and acts as a single point for interaction with the rest of the enterprise Grid. Aneka provides SLA support such that the user can specify QoS requirements such as deadline (maximum time period which the application needs to be completed in) and budget (maximum cost that the user is willing to pay for meeting the deadline). The user can access the Aneka Enterprise Grid remotely through the Gridbus broker. The Gridbus broker also enables the user to negotiate and agree upon the QoS requirements to be provided by the service provider.

## 4. Global Cloud Exchange and Markets

Enterprises currently employ Cloud services in order to improve the scalability of their services and to deal with bursts in resource demands. However, at present, service providers have inflexible pricing, generally limited to flat rates or tariffs based on usage thresholds, and consumers are restricted to offerings from a single provider at a time. Also, many providers have proprietary interfaces to their services thus restricting the ability of consumers to swap one provider for another.

For Cloud computing to mature, it is required that the services follow standard interfaces. This would enable services to be commoditised and thus, would pave the way for the creation of a market infrastructure for trading in services. An example of such a market system, modeled on real-world exchanges, is shown in Figure 4. The market directory allows participants to locate providers or consumers with the right offers. Auctioneers periodically clear bids and asks received from market participants. The banking system ensures that financial transactions pertaining to agreements between participants are carried out.

Brokers perform the same function in such a market as they do in real-world markets: they mediate between consumers and providers by buying capacity from the provider and sub-leasing these to the consumers. A broker can accept requests from many users who have a choice of submitting their requirements to different brokers. Consumers, brokers and providers are bound to their requirements and related compensations through SLAs. An SLA specifies the details of the service to be provided in terms of metrics agreed upon by all parties, and penalties for meeting and violating the expectations, respectively.

Such markets can bridge disparate Clouds allowing consumers to choose a provider that suits their requirements by either executing SLAs in advance or by buying capacity on the spot. Providers can use the markets in order to perform effective capacity planning. A provider is equipped with a price-setting mechanism which sets the current price for the resource based on market conditions, user demand, and current level of utilization of the resource. Pricing can be either fixed or variable depending on the market conditions. An admission-control mechanism at a provider's end selects the auctions to participate in or the brokers to negotiate with, based on an initial estimate of the utility. The negotiation process proceeds until an SLA is formed or the participants decide to break off. These mechanisms interface with the resource management systems of the provider in order to guarantee the allocation being offered or negotiated can be reclaimed, so that SLA violations do not occur. The resource management system also provides functionalities such as advance reservations that enable guaranteed provisioning of resource capacity.

Brokers gain their utility through the difference between the price paid by the consumers for gaining resource shares and that paid to the providers for leasing their resources. Therefore, a broker has to choose those users whose applications can provide it maximum utility. A broker interacts with resource providers and other brokers to gain or to trade resource shares. A broker is equipped with a negotiation module that is informed by the current conditions of the resources and the current demand to make its decisions.

Consumers have their own utility functions that cover factors such as deadlines, fidelity of results, and turnaround time of applications. They are also constrained by the amount of resources that they can request at any time, usually by a limited budget. Consumers also have their own limited IT infrastructure that is generally not completely exposed to the Internet. Therefore, a consumer participates in the utility market through a resource management proxy that selects a set of brokers based on their offerings. He then forms SLAs with the brokers that bind the latter to provide the guaranteed resources. The enterprise consumer then deploys his own environment on the leased resources or uses the provider's interfaces in order to scale his applications.

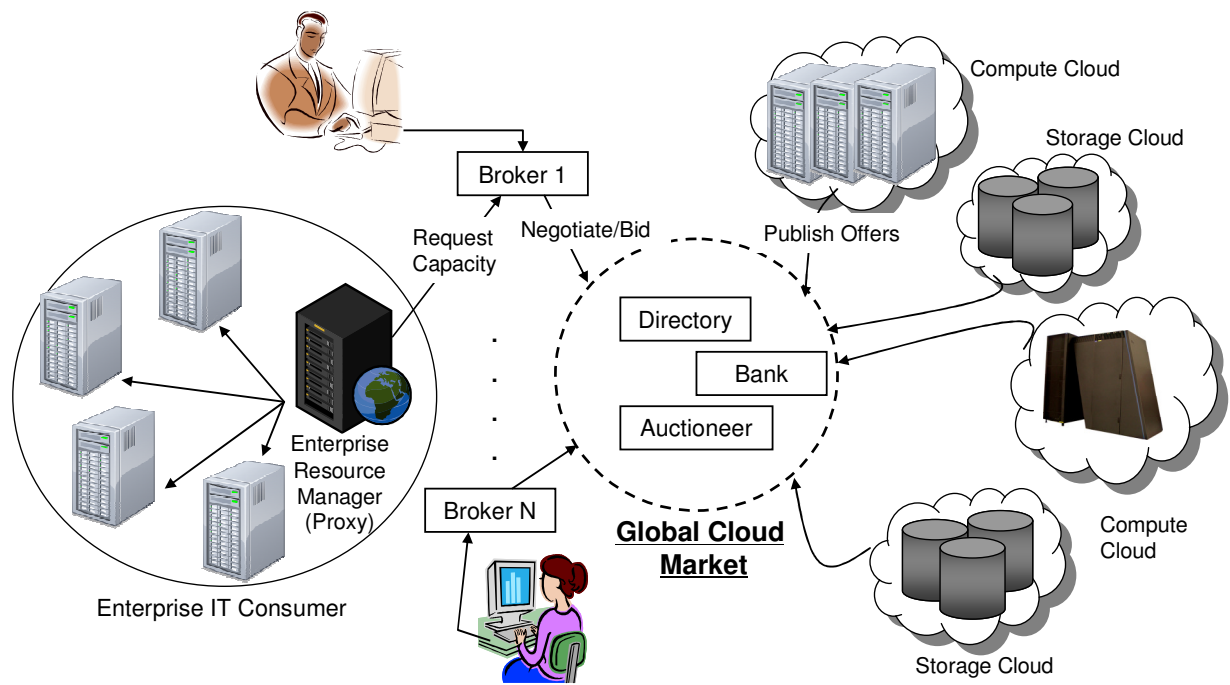

**Figure 4:** Global Cloud exchange and market infrastructure for trading services.

The idea of utility markets for computing resources has been around for a long time. Recently, many research projects such as SHARP [18], Tycoon [19], Bellagio [20], and Shirako [21] have come up with market structures for trading in resource allocations. These have particularly focused on trading in VM-based resource slices on networked infrastructures such as PlanetLab. As mentioned before, the Gridbus project has created a resource broker that is able to negotiate with resource providers. Thus, the technology for enabling utility markets is already present and ready to be deployed.

However, significant challenges persist in the universal application of such markets. Enterprises currently employ conservative IT strategies and are unwilling to shift from the traditional controlled environments. Cloud computing uptake has only recently begun and many systems are in the proof-of-concept stage. Regulatory pressures also mean that enterprises have to be careful about where their data gets processed, and therefore, are not able to employ Cloud services from an open market. This could be mitigated through SLAs that specify strict constraints on the location of the resources. However, another open issue is how the participants in such a market can obtain restitution in case an SLA is violated. This motivates the need for a legal framework for agreements in such markets, a research issue that is out of scope of themes pursued in this paper.

## 5. Summary and Conclusion

Cloud computing is a new and promising paradigm delivering IT services as computing utilities. As Clouds are designed to provide services to external users, providers need to be compensated for sharing their resources and capabilities. In this paper, we have proposed architecture for market-oriented allocation of resources within Clouds. We have discussed some representative platforms for Cloud computing covering the state-of-the-art. We have also presented a vision for the creation of global Cloud exchange for trading services.

The state-of-the-art Cloud technologies have limited support for market-oriented resource management and they need to be extended to support: negotiation of QoS between users and providers to establish SLAs; mechanisms and algorithms for allocation of VM resources to meet SLAs; and manage risks associated with the violation of SLAs. Furthermore, interaction protocols needs to be extended to support interoperability between different Cloud service providers.

As Cloud platforms become ubiquitous, we expect the need for internetworking them to create a market-oriented global Cloud exchange for trading services. Several challenges need to be addressed to realize this vision. They include: market-maker for bringing service providers and consumers; market registry for publishing and discovering Cloud service providers and their services; clearing house and brokers for mapping service requests to providers who can meet QoS expectations; and payment management and accounting infrastructure for trading services. Finally, we need to address regulatory and legal issues, which go beyond technical issues.

## Acknowledgements

This work is partially supported by the Australian Department of Innovation, Industry, Science and Research (DIISR) through International Science Linkage program.

## References

[1] L. Kleinrock. A vision for the Internet. *ST Journal of Research*, 2(1):4-5, Nov. 2005.

[2] S. London. INSIDE TRACK: The high-tech rebels. *Financial Times*, 06 Sept. 2002. http://search.ft.com/nonFtArticle?id=020906000371 [18 July 2008]

[3] I. Foster and C. Kesselman (eds). *The Grid: Blueprint for a Future Computing Infrastructure*. Morgan Kaufmann, San Francisco, USA, 1999.

[4] M. Chetty and R. Buyya. Weaving Computational Grids: How Analogous Are They with Electrical Grids? *Computing in Science and Engineering*, 4(4):61–71, July–Aug. 2002.

[5] A. Weiss. Computing in the Clouds. *netWorker*, 11(4):16-25, Dec. 2007.

[6] Twenty Experts Define Cloud Computing, http://cloudcomputing.sys-con.com/read/612375_p.htm [18 July 2008].

[7] R. Buyya, D. Abramson, and S. Venugopal. The Grid Economy. *Proceedings of the IEEE*, 93(3): 698-714, IEEE Press, USA, March 2005.

[8] S. Venugopal, X. Chu, and R. Buyya. A Negotiation Mechanism for Advance Resource Reservation using the Alternate Offers Protocol. In *Proceedings of the 16th International Workshop on Quality of Service (IWQoS 2008)*, Twente, The Netherlands, June 2008.

[9] D. Hamilton. 'Cloud computing' seen as next wave for technology investors. *Financial Post*, 04 June 2008. http://www.financialpost.com/money/story.html?id=562877 [18 July 2008]

[10] Morgan Stanley. Technology Trends. 12 June 2008. http://www.morganstanley.com/institutional/techresearch/pdfs/TechTrends062008.pdf [18 July 2008]

[11] K. Keahey, I. Foster, T. Freeman, and X. Zhang. Virtual workspaces: Achieving quality of service and quality of life in the Grid. *Scientific Programming*, 13(4):265-275, October 2005.

[12] I. Llorente, OpenNebula Project. http://www.opennebula.org/ [23 July 2008]

[13] Amazon Elastic Compute Cloud (EC2), http://www.amazon.com/ec2/ [18 July 2008]

[14] Google App Engine, http://appengine.google.com [18 July 2008]

[15] Microsoft Live Mesh, http://www.mesh.com [18 July 2008]

[16] Sun network.com (Sun Grid), http://www.network.com [18 July 2008]

[17] X. Chu, K. Nadiminti, C. Jin, S. Venugopal, and R. Buyya. Aneka: Next-Generation Enterprise Grid Platform for e-Science and e-Business Applications. In *Proceedings of the 3th IEEE International Conference on e-Science and Grid Computing (e-Science 2007)*, Bangalore, India, Dec. 2007.

[18] Y. Fu, J. Chase, B. Chun, S. Schwab, and A. Vahdat. SHARP: an architecture for secure resource peering. *ACM SIGOPS Operating Systems Review*, 37(5):133–148, Dec. 2003.

[19] K. Lai, L. Rasmusson, E. Adar, L. Zhang, and B. A. Huberman. Tycoon: An implementation of a distributed, market-based resource allocation system. *Multiagent and Grid Systems*, 1(3):169–182, 2005.

[20] A. AuYoung, B. Chun, A. Snoeren, and A. Vahdat. Resource allocation in federated distributed computing infrastructures. In *Proceedings of the 1st Workshop on Operating System and Architectural Support for the Ondemand IT Infrastructure (OASIS 2004)*, Boston, USA, Oct. 2004.

[21] D. E. Irwin, J. S. Chase, L. E. Grit, A. R. Yumerefendi, D. Becker, and K. Yocum. Sharing networked resources with brokered leases. In *Proceedings of the 2006 USENIX Annual Technical Conference (USENIX 2006)*, Boston, USA, June 2006.